\documentclass[pre,aps,twocolumn,superscriptaddress,letterpaper,showpacs]{revtex4}
\usepackage{amsmath}
\usepackage{amsfonts}
\usepackage{amssymb}
\usepackage{graphics}
\usepackage{graphicx}
\usepackage{color}
\usepackage[applemac]{inputenc}
\usepackage[cyr]{aeguill}
\usepackage{psfrag}

\newcommand{\ii}{\text{i}}
\newcommand{\ud}{\text{d}}

\newcommand{\um}{\mu\text{m}}
\newcommand{\e}{\text{e}}
\definecolor{green2}{rgb}{0,0.66,0} 

\begin{document}

\title{Time reversal sub-wavelength focusing in bubbly media}
\author{Maxime Lanoy}
\affiliation{Institut Langevin, ESPCI ParisTech, CNRS (UMR 7587), PSL Research University, Paris, France}
\affiliation{Laboratoire Mati\`ere et Syst\`emes Complexes, Universit\'e Paris-Diderot, CNRS (UMR 7057), Paris, France}

\author{Romain Pierrat}
\affiliation{Institut Langevin, ESPCI ParisTech, CNRS (UMR 7587), PSL Research University, Paris, France}

\author{Fabrice Lemoult}
\affiliation{Institut Langevin, ESPCI ParisTech, CNRS (UMR 7587), PSL Research University, Paris, France}

\author{Mathias Fink}
\affiliation{Institut Langevin, ESPCI ParisTech, CNRS (UMR 7587), PSL Research University, Paris, France}

\author{Valentin Leroy}
\affiliation{Laboratoire Mati\`ere et Syst\`emes Complexes, Universit\'e Paris-Diderot, CNRS (UMR 7057), Paris, France}

\author{Arnaud Tourin}
\affiliation{Institut Langevin, ESPCI ParisTech, CNRS (UMR 7587), PSL Research University, Paris, France}


\date{\today}

\begin{abstract}

Thanks to a Multiple Scattering Theory algorithm, we present a way to focus energy at the deep subwavelength scale, from the far-field, inside a cubic disordered bubble cloud by using broadband Time Reversal (TR). We show that the analytical calculation of an effective wavenumber performing the Independant Scattering Approximation (ISA) matches the numerical results for the focal extension. Subwavelength focusings of $\lambda/100$ are reported for simulations with perfect bubbles (no loss). A more realistic case, with viscous and thermal losses, allows us to obtain a $\lambda/14$ focal spot, with a low volume fraction of scatterers ($\Phi=10^{-2}$). Bubbly materials could open new perspective for acoustic actuation in the microfluidic context. 
\end{abstract}


\maketitle

\section{Introduction}
Time-reversal acoustics \cite{derode2001random, fink2008time} is an efficient way for focusing ultrasound deeply inside heterogeneous media.
In a typical time-reversal focusing experiment, the wave field radiated by a broadband source is recorded with a set of detectors located on a surface named a ``Time Reversal Mirror''. The detectors then become sources sending back the time-reversed sequence of the recorded field. The result is a time-reversed field that focuses at the source location. In the mid-nineties M. Fink and A. Derode carried out the first time reversal focusing experiment for ultrasound propagating in a disordered multiple scattering medium \cite{PhysRevLett.75.4206}. They showed that waves could thereby be focused with a finer resolution than in a homogeneous medium. That important result contributes to create a new paradigm for the manipulation of waves in complex media: contrary to long-held beliefs, disorder is not an impediment to focusing and imaging but can be turned into allies for controlling waves. 

More recently, it has been shown that it is even possible to beat the diffraction limit with the help of a complex medium structured on a subwavelength scale and using a time-reversal mirror placed in the far field \cite{Lerosey2007focusing}. This led to the concept of resonant metalens \cite{PhysRevLett.104.203901}, a new kind of lens comprising a dense arrangement of tiny resonators, each smaller than the relevant wavelength. The key idea is that a finite size medium made out of subwavelength resonators support modes that can radiate efficiently in the far field spatial information of the near field of a source.  First tested with microwaves, that concept has also been transposed to acoustics in the audible range \cite{PhysRevLett.107.064301} as well as in optics \cite{pierrat, lemoult2012polychromatic}. It actually appears that any medium of finite size consisting of resonant unit cells supports subwavelength modes and can therefore be used for subwavelength focusing from the far field using time reversal. Many of such media have been studied recently in the field of metamaterials, \textit{i.e.}, engineered materials consisting of a distribution (generally periodic) of sub-wavelength resonant unit cells.

Among all possible acoustic sub-wavelength resonators, bubbles in water seem to be very interesting candidates since  they exhibit a low-frequency acoustic resonance, known as the Minnaert resonance ~\cite{minnaert1933xvi}.  Here we investigate numerically the ability of an ultrasound time-reversal mirror to focus inside a bubbly metamaterial with a subwavelength resolution. To that goal we build a Multiple-Scattering Theory (MST) code that fully incorporates the multiple scattering effect. We carefully investigate the link between the wave dispersion observed in bubbly media and the spatial extension of the focusing spot obtained after the time reversal operation. We then address the most practicable case where viscous and thermal losses cannot be ignored by considering the appropriate damped scattering function. We highlight two regimes of frequencies that can be useful to achieve sub-wavelength focusing and examine carefully the influence of bubbles concentration.

\section{Multiple-Scattering Calculation}
Multiple scattering of acoustic waves is known to be strong in bubbly media, as shown in many numerical studies \cite{kafesaki, ye1998acoustic}. Especially they were presented by several authors as promising candidates for the study of Anderson localization \cite{sornette1988strong}. Multiple scattering calculation for a finite number of bubbles is simplified by the fact that air bubbles in a liquid can be considered as monopolar scatterers on a broad range of frequencies. Indeed, even at the first resonance of a bubble, known as the Minnaert\cite{minnaert} resonance, the wavelength in the liquid remains much larger than the size of the bubble. When excited by a monochromatic pressure $p^{(0)}\text{exp}[-\text{i}\omega t]$, with complex amplitude $p^{(0)}$, a bubble oscillates and generates at distance $r$ a spherical pressure field $p(r,t)=f^\text{s} p^{(0)} \text{exp}[\text{i} (k_0 r-\omega t)]/r$, where $k_0$ is the wavenumber in the liquid. The scattering function is given by 
\begin{eqnarray}
f^\text{s}&=& \frac{a}{(\omega_M/\omega)^2 -1-\text{i}\delta}, \label{eq1}
\end{eqnarray}
where $a$ is the radius of the bubble, $\omega_M$ its Minnaert resonance \cite{minnaert} and $\delta$ its damping constant \cite{devin1959survey,leighton2012acoustic,leroy2005bubble}.

\begin{figure*}[htb!]
    \centering
      \includegraphics[width=\linewidth]{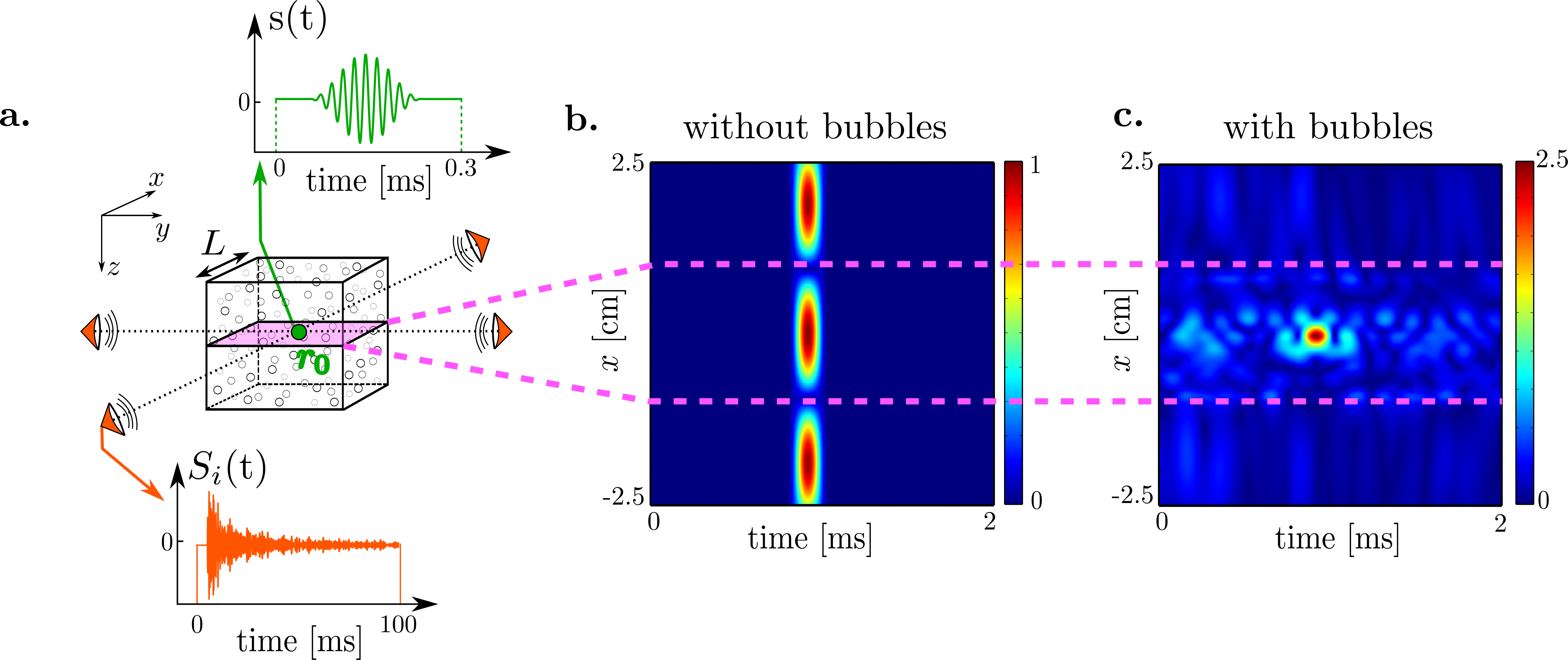}
    \caption{\textbf{a.} A source located at $\pmb{r_0}$ sends a broadband wave packet in the $50$-$60\,$kHz range and the field $S_i(t)$ is recorded by a 4-element TR mirror. \textbf{b.} Back-propagated field within a line in the $x$ direction as a function of time (b-scan), in pure liquid (no bubble). \textbf{c.} The same bscan when a $L=2\,$cm, $\Phi=10^{-4}$ cloud of $50\,\mu$m-radius bubbles is present.}\label{bscan}
\end{figure*}

In the case of a collection of $N$ bubbles, where bubble $i$ is described by its position $\pmb{r}_i$ and scattering function $f^\text{s}_i$, the pressure field experienced by bubble $i$ is not limited to the incident pressure $p^{(0)}_i$ but also includes the field generated by all the other bubbles \cite{lax1952multiple,leroy2008acoustic}:
\begin{equation}
p_i=p_i^{(0)}+\sum_{j\neq i}f^\text{s}_j p_j \frac{e^{\text{i}k_0\mid\mid\pmb{r}_j-\pmb{r}_i\mid\mid}}{\mid\mid\pmb{r}_j-\pmb{r}_i\mid\mid},
\label{p}
\end{equation}
where $c$  is the speed of sound in the liquid ($k_0=\omega/c$).
Eq.~(\ref{p}) can be written in a condensed matrix form:
\begin{equation*}
p_i = \sum_j M_{ij}^{-1} p^{(0)}_j \hspace{2pt} \text{with} \hspace{2pt}
M_{ij}=
     \begin{cases}
        \qquad 1 & \text{$i=j$} \\
        -4\pi f^\text{s}_j G_0(\omega; \pmb{r}_i,\pmb{r}_j)  &\text{$i\neq j$}
	      \end{cases},
\label{system}
\end{equation*}
 in which  $G_0(\omega; \pmb{r}_i,\pmb{r}_j)= \frac{\text{exp}[\text{i}k_0\mid\mid\pmb{r}_j-\pmb{r}_i\mid\mid]}{4\pi \mid\mid\pmb{r}_j-\pmb{r}_i\mid\mid}$ is the homogeneous Green's function between points $\pmb{r}_i$ and $\pmb{r}_j$.
The total pressure field at any point $\pmb{r} \neq \pmb{r}_i$ (\emph{i.e.} not occupied by a bubble) is the sum of the incident wave $p^0(\pmb{r})$ and the $N$ scattered waves. We thus obtain the effective Green function between $\pmb{r}_0$ and $\pmb{r}$:
\begin{multline}
G(\omega; \pmb{r}, \pmb{r}_0)=G_0(\omega; \pmb{r}, \pmb{r}_0)\\+ \sum_i   4\pi f^\text{s}_i G_0(\omega; \pmb{r},\pmb{r}_i)\Big[\sum_j M_{ij}^{-1} G_0(\omega; \pmb{r}_j, \pmb{r}_0) \Big].
\label{m}
\end{multline}
Providing that the locations and scattering functions of all the bubbles are known, Eq.~(\ref{m}) enables us to calculate the acoustic (linear) response of a bubble cloud to any excitation. It is important to note that the multiple scattering is fully taken into account by this scheme, without any approximations. Solving the multiple scattering problem thus amounts to inverting a $N\times N$ matrix, the only limitation being the number of scatterers $N$ included in the calculation. In practice, we were able to solve the MST with $N$ as large as $30,000$ bubbles due to computing ressources limitations.
\section{Time reversal numerical experiment}

In a typical time reversal (TR) experiment, a signal
$s(t) = \int p(\omega)\text{e}^{-\text{i}\omega t} \ud \omega$ is emitted by a point source at $\pmb{r}_0$, and recorded, after propagation through the multiple scattering medium, by receptors located at $\pmb{r}_i$, which acquire
\begin{equation}
S_i(t) = \int p(\omega)G(\omega ; \pmb{r}_i,\pmb{r}_0) \e^{-\ii \omega t} \ud \omega.
\end{equation}
The MST calculation is thus processed independantly for every frequences and the temporal signal is recovered performing an inverse Fourier Transform.
In figure~\ref{bscan}, we show the results of a numerical experiment performed with a $2\,$cm-edge cubic cloud corresponding to bubbles of radius $a=50\,\mu\text{m}$ in water, with a concentration of $210\,$ bubbles per cm$^3$  (gas volume fraction $\Phi=10^{-4}$). For this value of $a$, the resonance frequency of the bubbles is $\omega_M/(2\pi)=59\,$kHz, and we assume that the only source of damping for the bubbles is radiative $\delta_\text{rad}=k_0 a$, thus neglecting thermal and viscous dissipation at this stage. Also, the bubbles are assumed to remain still during the whole TR process. Although, this assumption could appear as a severe drawback for most bubbly liquids, it has been shown~\cite{leroy2009design, leroy2015superabsorption, leroy2009transmission} that it is possible to realize highly monodisperse and stable bubble collections thanks to yield-stress fluids or elastomers. A short wavepacket with a central frequency of $55\,$kHz is emitted from the center of the cloud, and four receptors record the signals at four different locations, in the far field. Note that, at such frequency, $\delta_\text{rad}=0.012$. When bubbles are present, the acquired signals are significantly longer than the emitted one, due to multiple scattering, as illustrated in Fig.~\ref{bscan}a. The last step of the TR consists in emitting the time-reversed signals $S_i(-t)$ by each of the receptors now acting as sources. Using Green function, one can easily calculate the back-propagated signal at point $\pmb{r}$:
\begin{equation}
P(\pmb{r},t) = \sum_i \int p^*(\omega) G^*(\omega; \pmb{r}_i, \pmb{r}_0) G(\omega; \pmb{r},\pmb{r}_i) \e^{-\ii \omega t} \ud \omega
\end{equation} 
where $^*$ denotes the complex conjugation.
Figure~\ref{bscan}b shows the time-reversed signal along the $x$-axis as a function of time (bscan), in the absence of bubbles. In such a homogeneous medium, TR produces a focused signal nearly as short as the initial one. However, the spatial focusing is limited by the diffraction limit. We obtain a focus spot with Full Width at Half Maximum (FWHM) that is slightly larger than the half-wavelength $\lambda/2$. Furthermore, intense side lobes are visible. As shown in Fig.~\ref{bscan}c, the presence of bubbles makes the focusing much more efficient: the focal spot is sub-wavelength in size, with no visible side lobes. To quantify the performance of the focusing, we compare the focal width to the diffraction limited width with a numerical aperture of 1, \emph{i.e.} $\lambda/2$. We thus look at dimensionless parameter $2\text{w}/\lambda$, where w is the measured FWHM and $\lambda$ the wavelength in the homogenous medium, at the central frequency. When focusing is limited by diffraction, one obtains $2\text{w}/\lambda\geq1$, whereas low values of $2\text{w}/\lambda$ indicate efficient sub-wavelength focusing. In Fig.~\ref{bscan}c, we measure w$=3.5\,$mm, which leads to $2\text{w}/\lambda=0.27$: the focal spot is almost four times smaller than the diffraction limit in free space. Note that this sub-wavelength focusing is obtained with a quite dilute bubbly medium: for $\Phi=10^{-4}$, the mean distance between two bubbles is larger than $30$ times their radii. 


\section{Prediction of the focal spot spatial extension}
The performance of the super-focusing depends on several parameters such as the frequency bandwidth, the size of the bubbles, and the number of bubbles per unit volume. In this section, we show it is possible to predict the focal spot length by analyzing the bubble cloud in terms of an effective medium. When a wave propagates through an heterogeneous medium, in the general case, the field splits into a coherent and a fluctuation contribution. It can be shown that the coherent term, which results from an averaging of the field over disorder, can be well described by an effective medium approach, i.e. by considering an homogeneous multiply scattering medium with an effective wavenumber $k$. The Independent Scattering Approximation (ISA) predicts that this effective wavenumber is given by 
\begin{equation}
k^2=k_0^2 + 4\pi n f^\text{s}, \label{ISA}
\end{equation}
where $k_0$ is the wavenumber in the host medium, $n$ the number of scatterers per unit volume, and $f^\text{s}$ their forward scattering function~\cite{foldy1945multiple}.

The ISA neglects loops of scattering, \textit{i.e.} sequences that involve the same scatterer several times. As the contribution of loops are expected to be strong at resonance in a bubbly medium~\cite{henyey1999corrections, ye1995acoustic}, corrections to the ISA should be taken into account. However, experiments with bubbly media up to $\Phi=10^{-2}$ showed that the prediction of the ISA was in good agreement with the experimental data, even at resonance~\cite{leroy2008acoustic}. 

\begin{figure}[htb!]
    \centering
      \includegraphics[width=\linewidth]{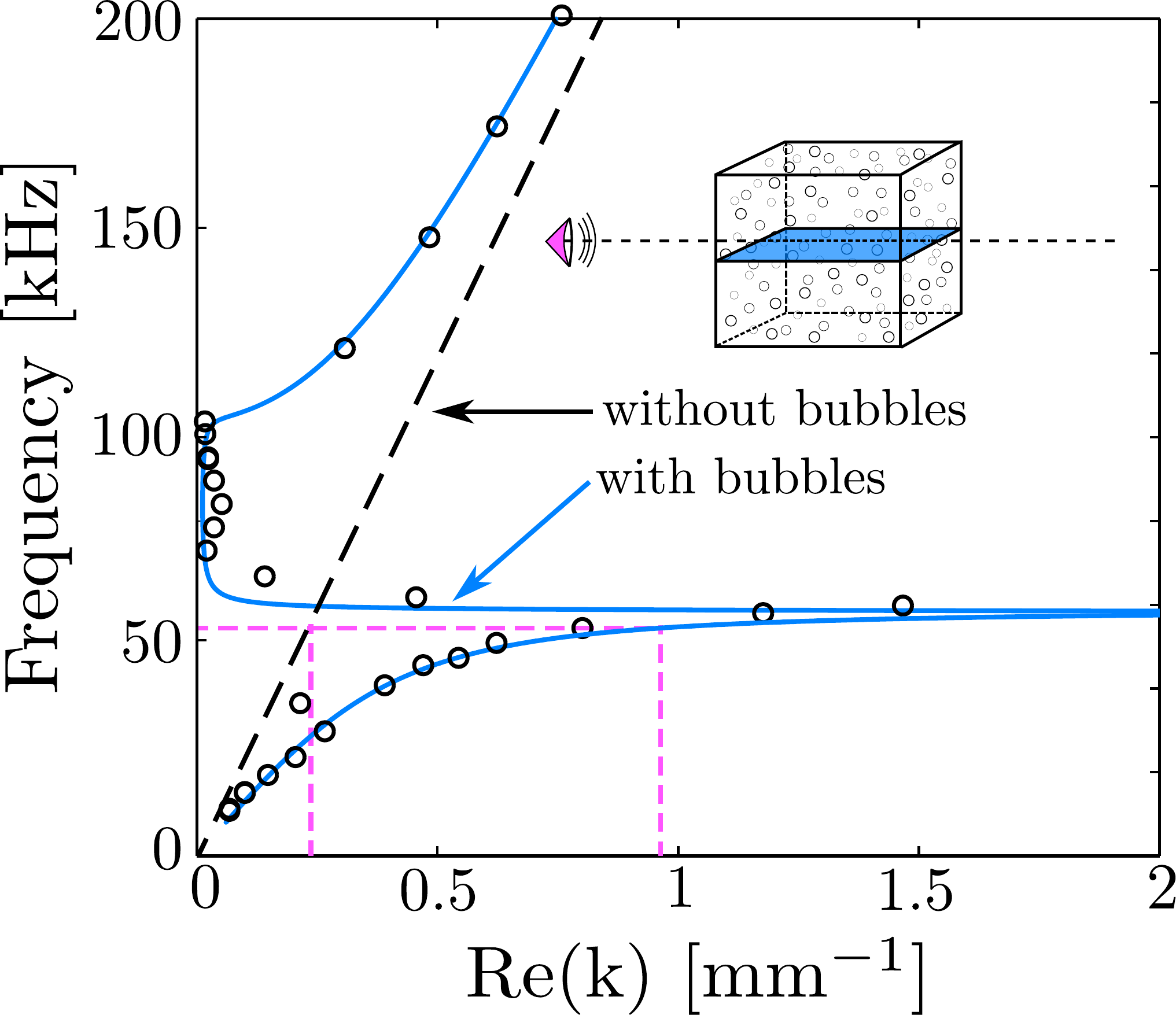}
    \caption{Dispersion relation for a $\Phi= 10^{-4}$ bubbly medium with $50\,\um$-radius bubbles. Lines are ISA prediction for the real part of the wavenumber (blue solid line for the bubbly medium, black dashed line for free space), symbols come from a numerical experiment with the MST code. The numerical experiment consists in sending a monochromatic signal on a large cloud of bubbles ($L=7\,$cm), recording the field within an horizontal section of the cube (see inset), and analyzing this field to measure the effective wavenumber. Only the real part of the wavenumber is shown here, but the imaginary part was also checked to match ISA prediction.}\label{Disp}
\end{figure}

\begin{figure*}[htb!]
    \centering
      \includegraphics[width=\linewidth]{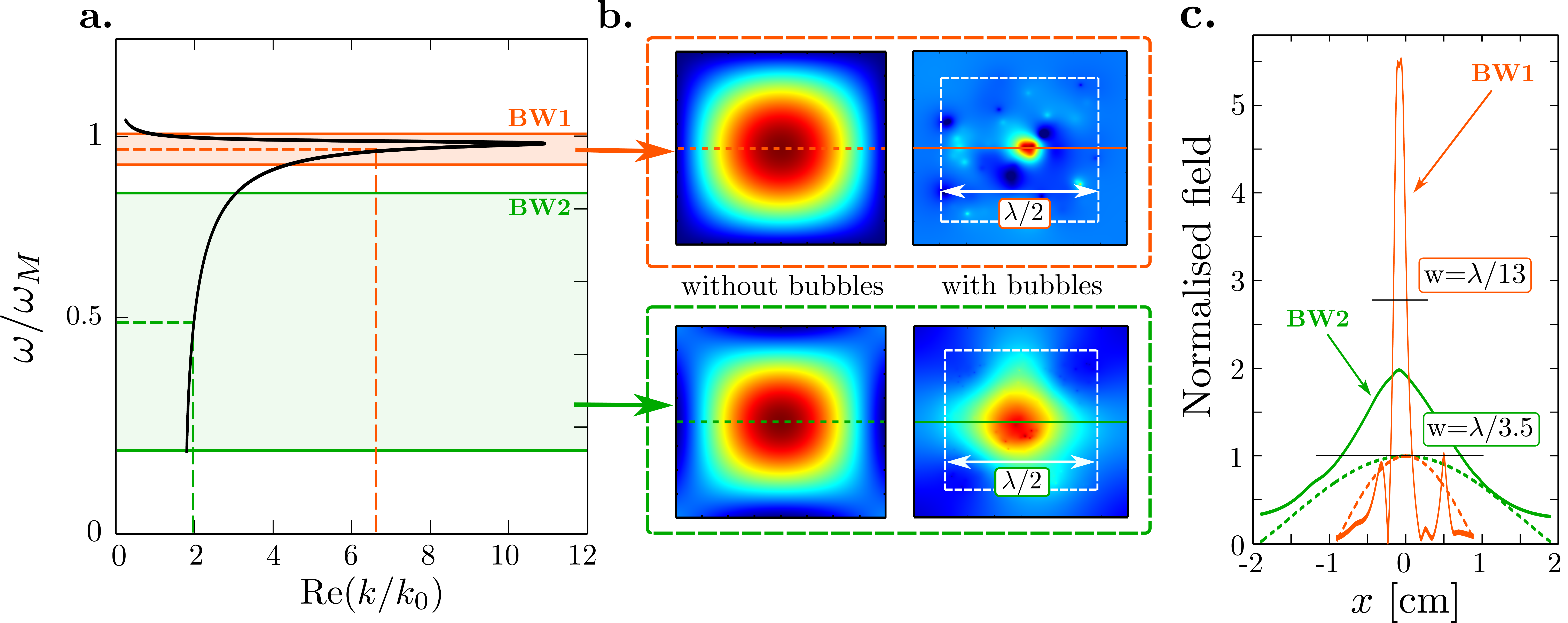}
    \caption{\textbf{a.} The same dispersion relation as in Fig~\ref{Disp}, rescaled by the resonance angular frequency $\omega_M$ and the wavenumber in liquid $k_0$. Two bandwidths are highlighted: BW1 is close to the resonance ($56$-$59\,$kHz, in red), with an average $k/k_0$ of about $6$; BW2 is on the low frequency branch ($10$-$45\,$kHz, in green), with $k/k_0$ around $2$. \textbf{b.} Pressure fields obtained on the central section of a $\Phi=10^{-4}$ cubic cloud when TR focusing is performed on BW1 (top) and BW2 (bottom),  in the absence (left) and with scatterers (right). Bubble clouds sizes are $\lambda/2$, i.e. adapted to the central wavelength of the chosen bandwidth ($\lambda=2.6\,$cm on BW1, centered at $57.5\,$kHz; and $\lambda=5.5\,$cm on BW2, centered at $27.5\,$kHz). \textbf{c.} Pressure fields measured within the colored line of b and normalized by the the field obtained without bubble. We measure $\text{w}=2\,$mm for BW1 and $\text{w}=1.5\,$cm for BW2.}\label{RT}
\end{figure*}

Fig.~\ref{Disp} shows the dispersion relation predicted by the ISA for bubbly water with a gas volume fraction $\Phi=10^{-4}$  and perfect bubbles (without dissipation) of radius $a=50\,\um$. The bubbly medium appears as very dispersive, with three regimes. (i) Below $59\,$kHz (i.e. the Minnaert resonance of the bubbles), the wavenumber is purely real and corresponds to a phase velocity ($\omega/k$) that is below that in pure water (black dashed line in Fig.~\ref{Disp}). (ii) Between $59$ and $100\,$kHz, waves are evanescent, with very low values of the real part of $k$. (iii) Above $100\,$kHz, the waves become propagating again, the wavenumber getting closer to the value for propagation in pure water as frequency increases. 
These three regimes of propagation can be retrieved with our MST calculation, as shown by solid symbols in Fig.~\ref{Disp}.
The typical S-shape of the dispersion relation is the signature of a polariton-like dispersion (aka hybridization), due to the presence of low-frequency resonators~\cite{lagendijk1993vibrational}. From the focusing point of view, the low frequency branch is particularly interesting because the effective wave propagates with wavenumbers significantly larger than the wavenumber in pure water (compare the solid and dashed curves at low frequencies in Fig.~\ref{Disp}). It means that, thanks to bubbles, the waves can carry spatial information that would have been lost in pure water. It is then tempting to predict that the FWHM of the focal spot is directly linked to the effective wavelength on the considered bandwidth. For instance, the $\text{w}=3.5\,$mm focal spot obtained in Fig.~\ref{bscan} could be interpreted as coming from a $7\,$mm effective wavelength, \emph{i.e.} a wavenumber $2\pi/7=0.9\,$mm$^{-1}$ at $55\,$kHz, which is indeed well verified in Fig.~\ref{Disp} (see dashed-lines). 
\begin{figure}
    \centering
      \includegraphics[width=\linewidth]{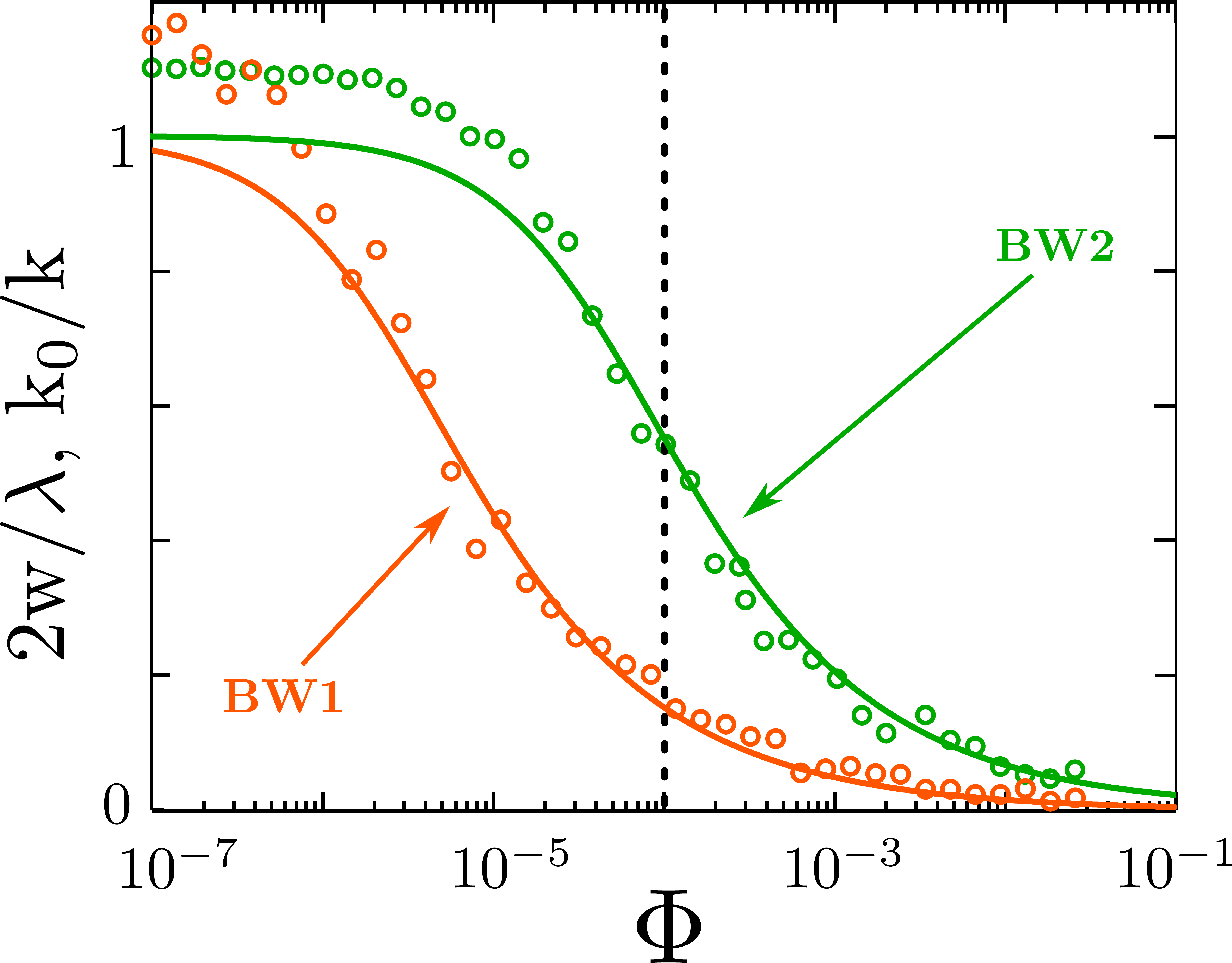}
    \caption{Dimensionless FWHM $2$w$/\lambda$ as a function of gas volume fraction $\Phi$ (symbols), compared to the dimensionless effective wavelength $k_0/k$ predicted by the ISA (lines).}\label{frac}
\end{figure}

To test this hypothesis, we have carried out two numerical TR experiments with the same bubble cloud as previously, but on different frequency bandwidths (BW). 
BW1 is close to the resonance of the bubbles, \emph{i.e.} at frequencies for which the ISA predicts a high effective wavenumber, with $k/k_0\simeq 6$ (see Fig.~\ref{RT}a). BW2 corresponds to the low frequency part of the dispersion, with $k/k_0\simeq 2$, almost constant on the bandwidth. The results, reported in Fig.~\ref{RT}, show that in both cases the RT operation successfully focus at the targeted position. Figs~\ref{RT}b and c confirm that focusing with BW1 beats the diffraction limit by a factor $6$, whereas a factor $2$ is obtained with BW2. A further confirmation of the effective medium interpretation is brought by Fig.~\ref{frac}: the FWHM of the focal spot is measured for increasing gas volume fractions $\Phi$ and compared to the ISA prediction on the two bandwidths. Surprisingly, even on the resonant BW, the ISA is found to correctly predict the width of the focal spot. We also emphasize that monochromatic focusing cannot yield to such performances and that the use of broadband TR is required.
\begin{figure}
        \centering
      \includegraphics[width=\linewidth]{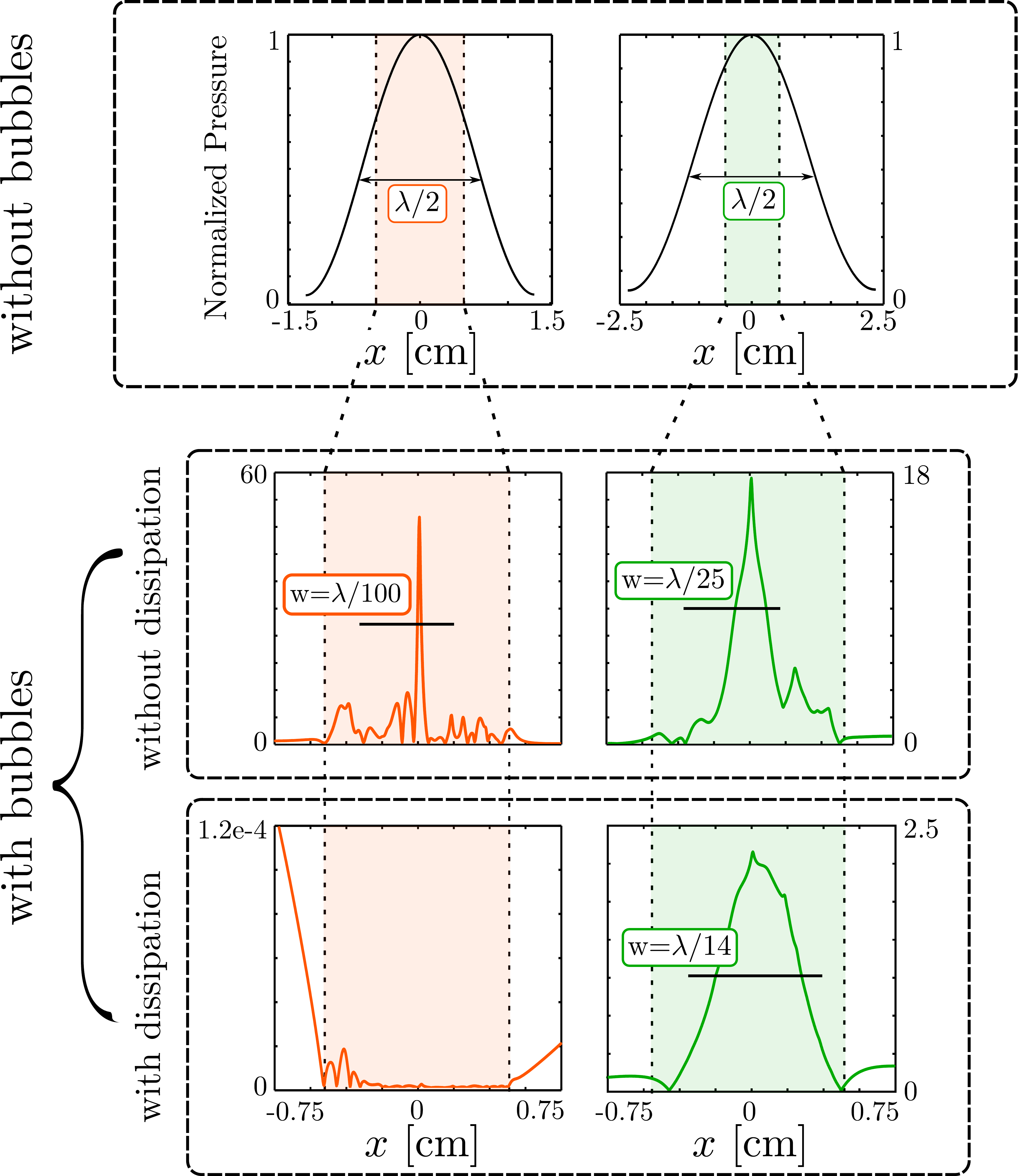}
        \caption{Focal spots obtained with BW1 (left) and BW2 (right) in three configurations: in pure water (up), in a $\Phi=0.01$ bubble cloud with no dissipation (middle), and in the same cloud with dissipation (bottom). The bubble cloud is a $1$-cm-edge cube. Fields are normalized by the maximum of pressure obtained in the absence of bubbles.}
\label{rt2}
\end{figure}

\section{Accounting for dissipation}
As evidenced in Fig~\ref{frac}, very efficient subwavelength focusing can be obtained with perfect bubbles. For instance with $\Phi=10^{-2}$, a TR focusing on BW1 can lead to a $\lambda/100$ spot. However, such focal spots can only be obtained with numerical experiments, in which dissipation can be totally neglected. In practice, bubbles are strong scatterers but they also absorb part of the incoming energy, due to viscous and thermal losses. Thermal losses are proportional to $\gamma -1$ \cite{prosperetti1977thermal}, $\gamma$ being the ratio of the heat capacities of the gas. Viscous losses are proportional to $\eta$, the viscosity of the liquid. Let us examine a realistic case with $\gamma=1.1$ (air with vapors of perfluorohexane, for instance), and $\eta=10^{-3}\,$Pa.s (water). With such values, the damping constant of a single bubble, which appears in Eq.~(\ref{eq1}), is $\delta = 0.04$, while it was $\delta = 0.012$ in previous simulations, when only radiative damping was considered. 

\begin{figure}
    \centering
      \includegraphics[width=\linewidth]{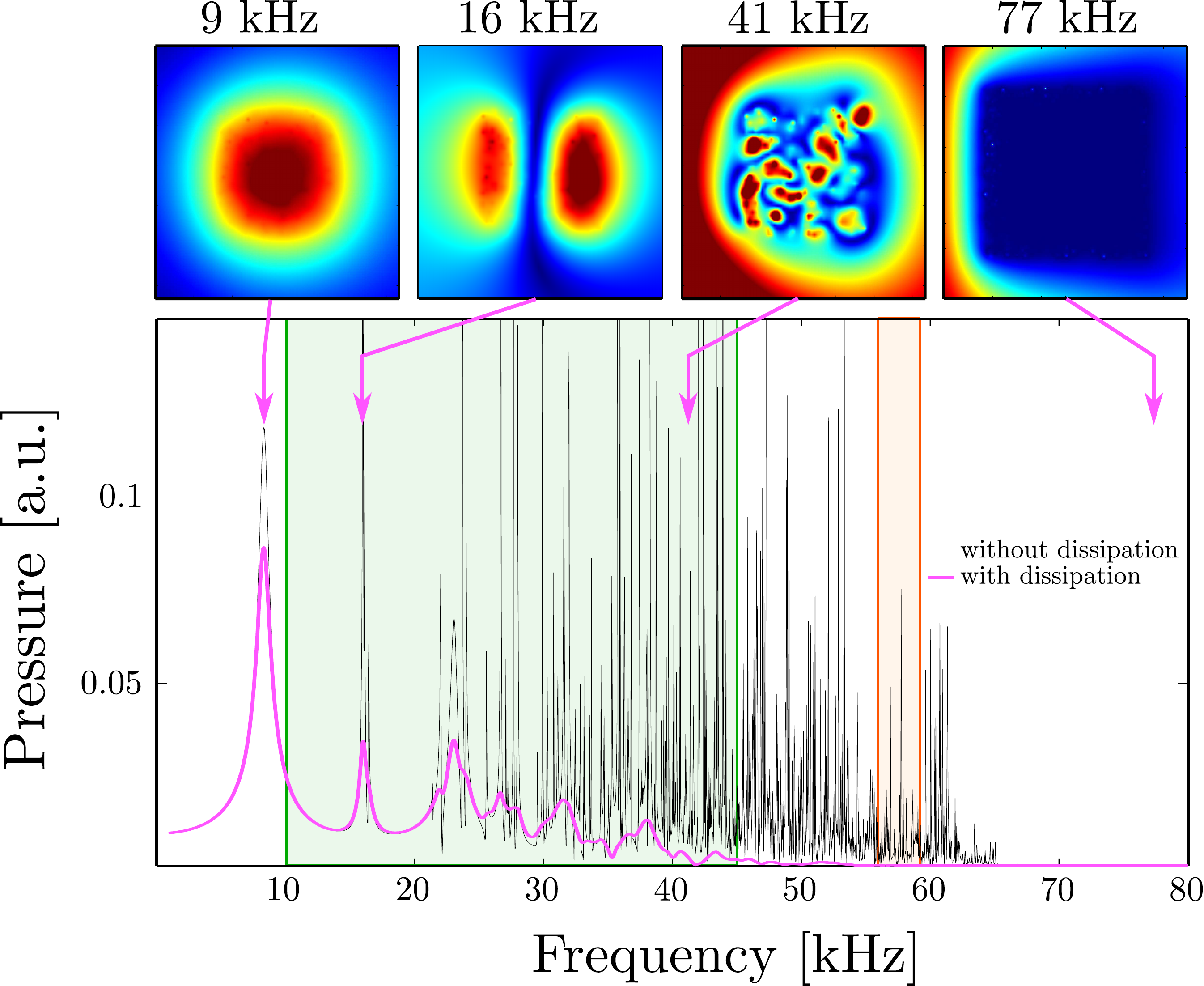}
    \caption{Pressure spectra obtained at one point of the $z=0$ section of a $\Phi=0.01$ bubble cloud with (thick pink line) and without (thin black line) dissipation. Edges of the cubic cloud are $1\,$cm long. Insets on the top are 2D field maps at different frequencies (represented by pink arrows) within the $z=0$ section, in the case of bubbles with dissipation.}
\label{modes}
\end{figure}

Figure~\ref{rt2} reports the results of the TR numerical experiments in a $\Phi=10^{-2}$ bubble cloud without and with dissipation, on the two bandwidths. As already seen in Fig.~\ref{frac}, extremely narrow focal spots are obtained when no dissipation is considered: $\lambda/100$ on BW1, and $\lambda/25$ on BW2. When dissipation is taken into account, the situation is much different: no focal spot is recovered on BW1, and the one on BW2 broadens but reamins sub-wavelength with a width of $\lambda/14$. The effect of dissipation is thus stronger for frequencies close to the resonance of the bubbles. Yet, even with dissipation, the S-shape of Fig.~\ref{Disp} is preserved, and the ISA still predicts very high values for the real part of the effective wavenumber. The explanation comes from an ingredient we have neglected so far: modes. Having access to high wavenumbers is not sufficient for subwavelength focusing, one also needs to excite enough modes, with the right phase, to build the desired spot. The effect of dissipation on modes is well illustrated in Fig.~\ref{modes}, which shows the pressure measured at one point of the bubble cloud as a function of frequency, for two cases: with and without dissipation. Without dissipation, many narrow peaks are visible in the spectrum. They are signatures of many modes implying individual scatterers. As the density of modes is high, a quite narrow BW is sufficient to gather many of them and thus reconstruct the focusing spot. The exact combination of modes one needs to reconstruct the desired spot is provided by the TR process. When dissipation occurs, the quality factors decrease and modes start to overlap. As a consequence, fewer modes can be controlled independently and the focus spot deteriorates \cite{lemoult2011far}. This effect appears to be even more critical for high frequencies. Indeed, the closer we get to the Minnaert resonance, the smaller the group velocity becomes. As a consequence, corresponding modes have longer life times in the medium and suffer more from dissipation. On BW1, all the modes disappear because of dissipation, making subwavelength focusing ineffective. On BW2, on the other hand, the effect of dissipation is less dramatic and some of the modes survive to dissipation. Examples of modes are represented in Fig.~\ref{modes} (see insets). Because less modes are available than in the ideal case, the w$=\lambda/25$ predicted by the ISA cannot be obtained, but a good $\lambda/14$ subwavelength focusing is still possible (see Fig.~\ref{rt2}).

It thus turns out that subwavelength focusing with bubbles can be very effective, but not for frequencies close to the individual resonance of the bubbles because, in realistic cases, the quality factor of the bubbles is not good enough to ensure the presence of independent modes available for the reconstruction of the spot. It is well-known that a good quality factor is indeed a critical requisite for subwavelength focusing with resonators~\cite{PhysRevLett.104.203901,lemoult2011far}. Another limitation of focusing at resonance is the sensitivity to polydispersity: numerical investigations (not shown here) with polydisperse bubble clouds showed that a 5$\%$ polydispersity severely impacts the quality of the focal spot on BW1. In bubbly media, subwavelength focusing is nevertheless possible, at low frequencies. Indeed, as shown in Fig.~\ref{Disp} even for frequencies much lower than the resonance frequency, the dispersion relation in a bubbly medium gives access to high wavenumbers. It comes from the high compressibility of bubbles, which makes bubbly liquids \emph{compressible} liquids, with a particularly low effective phase velocity at low frequencies, and hence a small effective wavelength~\cite{wood1977effective}. Further more, the focusing was found to remain unaffected by polydispersity on this range.

\section{Conclusion}
Broadband Time Reversal in perfect 3D bubbly media leads to very efficient subwavelength focusing. For instance, a bubble volume fraction of $\Phi = 10^{-2}$ can lead up to a super-resolution of $\lambda/100$. It was also evidenced that the spatial extension of the spot matches the effective wavelength predicted by the Independent Scattering Approximation (see Fig.~\ref{frac}). We then considered the realistic case where thermal and viscous losses are included. Although the focusing becomes totally ineffective in the resonant range, we take advantage of the compressibility of bubbly liquids in order to reach low effective wavelengths in the low frequency regime where the modes remain usable. Under such conditions we were able to reach a resolution of $\lambda/14$. This result could still benefit from optimization techniques such as Inverse Filter \cite{tanter2000time} or Iterative Time Reversal \cite{montaldo2004real} to enhance the resolution.

One of the perspectives of this work is to use bubbly materials to focus a large amount of acoustic energy on a small volume. As shown in Fig.~\ref{rt2}, the presence of the bubbles does not only make the spot narrower, it also multiply the amplitude of the field by more than a factor two. In practice, one could imagine to deposit even more energy, thanks to the time compression offered by the TR operation \cite{fink2003time}. An important aspect of bubbly media is the low concentration of bubbles one needs to obtain strong effects.  For $\Phi=0.01$, the bubbles are $7.5$ radii apart, which means there is space to consider placing other objects in the medium. This could open up new horizons for manipulation of objects in microfluidic flows, for example~\cite{RevModPhys.83.647, huerre2014bubbles}. Many microfluidic devices are actually fabricated in a soft elastic solid, in which air cavities behave as bubbles~\cite{leroy2009design}, making the calculations described in this article directly applicable. 

\section{Acknowledgements}
This work is supported by LABEX WIFI  (Laboratory of Excellence within the French Program ``Investments for the Future'') under references ANR-10-LABX-24 and ANR-10-IDEX-0001-02 PSL*. We thank Direction G\'en\'erale de l'Armement (DGA) for financial support to M.L.

\end{document}